\documentclass[%
reprint,
aip,
amsmath,amssymb,
aps,
]{revtex4-2}
\usepackage{graphicx}% Include figure files
\usepackage{dcolumn}% Align table columns on decimal point
\usepackage{bm}% bold math

\newcommand{\jgr}{J. Geophys. Res.}
\newcommand{\apjl}{Astrophys. J. Lett.}
\newcommand{\grl}{Geophys. Res. Lett.}
\newcommand{\araa}{Ann. Rev. Astron. Astrophys.}

\begin{document}
\preprint{APS/123-QED}
\title{Nonlinear Explosive Magnetic Reconnection in a Collisionless System} 
\author{Masahiro Hoshino}
\affiliation{Department of Earth and Planetary Science, The University of Tokyo, Tokyo 113-0033, Japan.}
\email{hoshino@eps.s.u-tokyo.ac.jp}
\begin{abstract}
The debate surrounding fast magnetic energy dissipation by magnetic reconnection has remained a fundamental topic in the plasma universe, not only in the Earth's magnetosphere but in astrophysical objects such as pulsar magnetospheres and magnetars, for more than half a century.
Recently, nonthermal particle acceleration and plasma heating during reconnection have been extensively studied, and it has been argued that rapid energy dissipation can occur for a collisionless ``thin'' current sheet, the thickness of which is of the order of the particle gyro-radius.  However, it is an intriguing enigma as to how the fast energy dissipation can occur for a ``thick'' current sheet with thickness larger than the particle gyro-radius.  Here we demonstrate, using a high-resolution particle-in-cell simulation for a pair plasma, that an explosive reconnection can emerge with the enhancement of the inertia resistivity due to the magnetization of the meandering particles by the reconnecting magnetic field and the shrinkage of the current sheet.  In addition, regardless of the initial thickness of the current sheet, the time scale of the nonlinear explosive reconnection is tens of the Alfv\'{e}n transit time. 
\end{abstract}
\maketitle
%\tableofcontents
%--------------------------------------------------------------------------
\section{Introduction}
%--------------------------------------------------------------------------
Recently, magnetic reconnection has received great attention in various plasma environments of the Earth's magnetosphere, laboratory plasmas, and astrophysical objects such as pulsar magnetospheres and magnetars \citep[e.g.,][]{Birn07,Zweibel09,Hoshino12,Uzdensky16,Blandford17}.  In these hot and dilute magneto-active plasmas, the magnetic energy stored in a plasma sheet with an anti-parallel magnetic field component can be efficiently converted into not only the bulk flow energy but also the energies of plasma heating and nonthermal particle acceleration. The reconnection paradigm has been established as a powerful magnetic energy dissipation process in the plasma universe \cite[e.g.,][]{Coppi66,Hoh66,Schindler66,Biskamp71,Galeev76}.  
In addition, in the last couple of decades, our understanding of reconnection dynamics has gradually accumulated for a ``thin'' plasma sheet with a thickness of the order of the particle gyro-radius, and it is recognized that fast energy dissipation can occur for such a thin plasma sheet. However, the mechanism of fast energy dissipation of reconnection for a ``thick'' plasma sheet has remained an intriguing enigma for more than half a century.  According to the linear theory of collisionless tearing modes, the growth rate $\gamma_{\rm l} \tau_{\rm A}$ 
is known to be proportional to $(r_{\rm g}/\lambda)^{3/2}$, where $r_{\rm g}$, $\lambda$, and $\tau_A$ are the gyro-radius of the thermal particle, the thickness of the current sheet, and the Alfv\'{e}n transit time defined by $\lambda/v_A$, respectively.  The linear theory suggests that the energy dissipation for a thick plasma sheet with $r_{\rm g} \ll \lambda$ is extremely slow.

To overcome the difficulty of the fast energy dissipation for a thick plasma sheet, many people have challenged the mechanisms of the explosive nonlinear growth of the reconnecting magnetic flux in the fluid approximation so far \cite[e.g.,][]{Ottaviani93,Cafaro98,Shay04,Bhattacharjee05,Comisso13,Hirota15}.
Although the standard resistive magnetohydrodynamics (MHD) model was known not to initiate a fast reconnection unless an anomalous resistivity is locally enhanced around the X-type neutral point \cite[e.g.][]{Biskamp86,Kulsrud01},
two-fluid/collisionless effects such as electron inertia, Hall current, ion gyro-radius have been argued to drive a fast reconnection.
During the early time evolution, a rapid shrinkage of the current sheet can happen, and the thinning of the current sheet into such kinetic scales can lead to the explosive reconnection.  It has been argued that the topological conservation of the Lagrangian invariant leads to the formation of the thin current sheet in collisionless reconnection \Citep{Cafaro98,Comisso13}.

In addition to the shrinkage of the current sheet in collisionless reconnection, \citet{Galeev78} proposed another idea of the explosive reconnection mechanism by focusing on the time evolution of the collisionless inertia resistivity in association with the growth of the reconnecting magnetic field.  They considered the enhancement of the inertia resistivity due to magnetized particles by the reconnecting magnetic field around the X-type neutral line,
and estimated that the nonlinear growth rate $\gamma_{\rm G} \tau_A$ was of the order of $(r_{\rm g}/\lambda)^{1/2} b_1$, where $b_1=B_1/B_0$ is the ratio of the reconnection magnetic field $B_1$ to the magnetic field $B_0$ outside of the plasma sheet.  This behavior has been demonstrated by particle-in-cell (PIC) simulation \cite{Terasawa81}.
This nonlinear reconnection suggests not only much faster energy dissipation than in the linear stage, but also an explosive nature of growth, namely, $b_1 (t)=b_1(t_0)/(1 - (t-t_0)\gamma_{\rm l})$, where $t_0$ is the start time of the nonlinear reconnection.  
The nonlinear explosive growth, however, still depends on the initial plasma thickness of $r_{\rm g}/\lambda$, and the time scale is of the order of $\gamma_{\rm l}^{-1}$.  Therefore, the fast reconnection issue for a thick current sheet remains unresolved.

In this paper, based on high-resolution two-dimensional PIC simulation, we propose a new mechanism of nonlinear reconnection in which the energy dissipation can occur in the time scale of the Alfv\'{e}n transit time.  The crucial factor is the enhancement of the inertia resistivity due to the effects of both thinning of the current sheet and the reconnecting magnetic field during reconnection.  As the key process, we show that the current sheet shrinks quickly to a thickness of the gyro-radius in PIC simulations for a pair plasma.  The theoretical model to explain the nonlinear explosive reconnection is also discussed.
%--------------------------------------------------------------------------
\section{Overview of Simulation Study}
%--------------------------------------------------------------------------
Let us first study the time evolution of the tearing-mode instability using two-dimensional PIC simulation \cite{Hoshino87,Hoshino98,Hoshino18} with a periodic boundary in the $x$-direction and the conducting walls for the upper and lower boundaries at $y= \pm 8 \lambda$, where $\lambda$ is the thickness of the initial plasma sheet. The total system size $L_x \times L_y$ is $108 \lambda \times 16 \lambda$ and the computational grid size is $5376 \times 800$. The total number of particles is $1.2 \times 10^{11}$ to calculate the high-$\beta$ plasma sheet as accurately as possible.
This large number of particles is necessary for the linear growth of the tearing mode to overcome thermal fluctuations, the amplitudes of which are proportional to $1/\sqrt{n_{\rm p}}$, where $n_{\rm p}$ is the number of particles per grid cell.  

For simplicity, we adopt the Harris solution \citep{Harris62} for the pair plasma with the same mass $m$ and temperature $T$, and we focus on a ubiquitous nonlinear reconnection mechanism in a collisionless plasma system.  The plasma temperature $T=mc^2/10$ and the Alfv\'{e}n speed $v_{\rm A}=B_0/\sqrt{8 \pi m N_0}=0.447 c$ are maintained for all simulation runs.  The thermal velocity $v_{\rm th}$ is equal to $v_{\rm A}$.
The magnetic field $B$ and plasma density $N$ are given by $B_x(y) = B_0 \tanh( y/\lambda)$ and $N(y)=N_0 \cosh^{-2}(y/\lambda) + N_{\rm b}$, 
respectively, where $N_{\rm b}$ is the background uniform plasma density. The background plasma density $N_{\rm b}$ is set to be $5 \%$ of the maximum Harris plasma density $N_0$. In addition, we have the relationships of the pressure balance $B_0^2/ 8 \pi = 2 N_0 T$ and the force balance $2 T/\lambda = |e| \beta B_0$, where $\beta c$ is the drift velocity.  Three different cases of $\beta=0.05$, $0.1$, and $0.2$, which correspond to $r_{\rm g}/\lambda=0.11$, $0.22$, and $0.45$, are studied.  Note that to date most simulation studies have investigated a thin plasma sheet such as $r_{\rm g}/\lambda \sim 0.5 - 2.0$. \cite[e.g.,][]{Hesse01,Pritchett01,Hoshino01,Drake08}

We add a small initial perturbation for the vector potential $\delta A_z$ to the Harris equilibrium, which is given by 
\begin{equation*}
  \delta A_z(x,y) \propto \exp (-(\frac{x-x_0}{\lambda_x})^2-(\frac{y}{\lambda_y})^2 )
  \cosh (\frac{y}{\lambda_y}),
\end{equation*}
where $\lambda_x = 2 \lambda$, $\lambda_y = \lambda /2$, $x_0=54 \lambda$, and the initial amplitude of the reconnected magnetic field is set to be $\max(B_y/B_0)=10^{-3}$ in the neutral sheet.
The reason of the initial perturbation is to place the X-type point in the center of the simulation box and to save the computational CPU time.
We have also studied the case without $\delta A_z(x,y)$, and observed the almost same time evolution as the case with $\delta A_z(x,y)$.  Other remarks will be given in the section of Discussion and Summary.
Although this perturbation mimics an X-type reconnection structure, the initial perturbation does not necessarily satisfy an eigenfunction for the linear tearing mode. Therefor, broadband electromagnetic waves are generated. To suppress these waves, a numerical filter outside the plasma sheet is used during the early phase with $t/\tau_{\rm A} < 16$.

%----------------------------------------------------------------------------
\begin{figure}
\includegraphics[width=8cm]{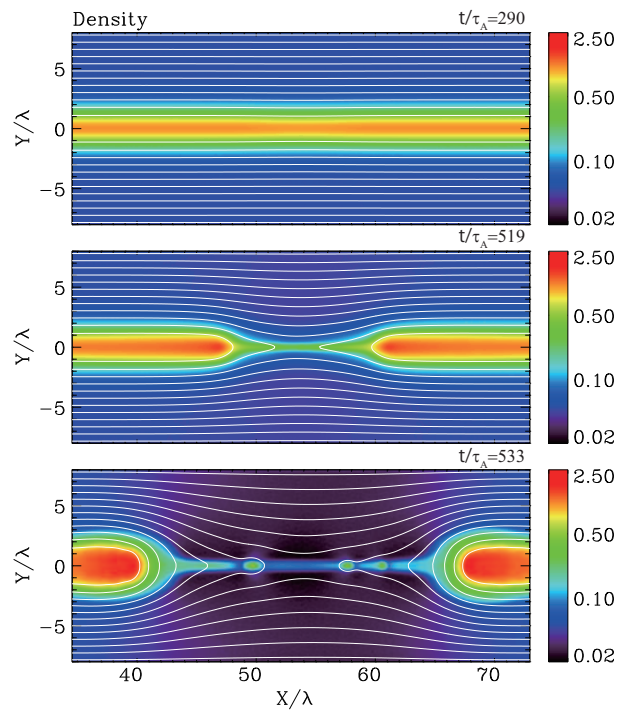}
\caption{Time evolution of the collisionless reconnection for a thick plasma sheet with $r_{\rm g}/\lambda=0.11$ obtained by PIC simulation.  From top to bottom are the early linear stage when the flux of the reconnecting magnetic fields is very small ($t/ \tau_{\rm A}=290$), the nonlinear stage with thinning of the plasma sheet around the X-type neutral line ($t/ \tau_{\rm A}=519$), and the late nonlinear stage where several small plasmoids are formed in the elongated plasma sheet ($t/ \tau_{\rm A}=533$).}
\label{fig:FIG1}
\end{figure}
%----------------------------------------------------------------------------

%----------------------------------------------------------------------------
\begin{figure*}
\includegraphics[width=18cm]{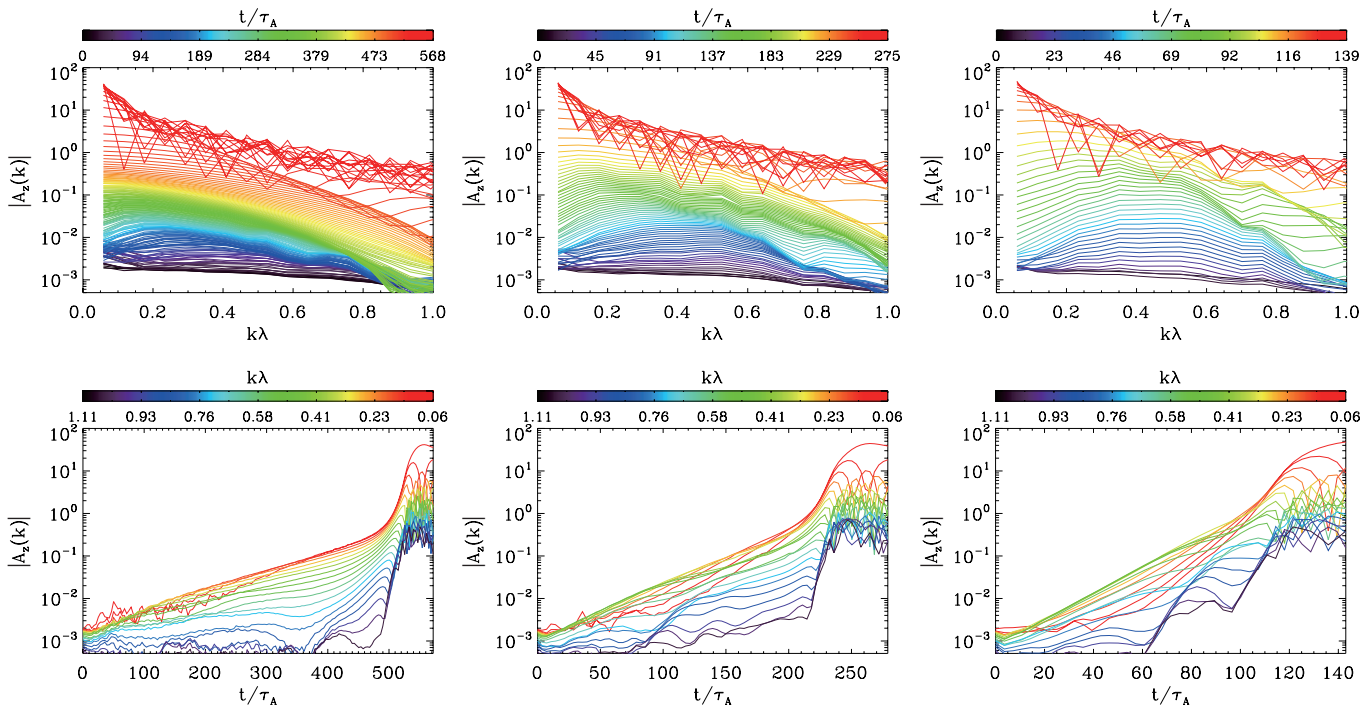}
\caption{Time evolution of the Fourier power spectra of the vector potential $|A_z|$ for three cases of $r_g/\lambda=0.11$ (left), $0.22$ (middle), and $0.45$ (right). The top three panels are the evolution of $|A_z|$ as the function of the wave number $k_x \lambda$, and the bottom three panels are those as the function of time $t/\tau_{\rm A}$.  Both are depicted from the same data set.  The colors on the top and the bottom represent the time stage of $t/\tau_{\rm A}$ and the wave number $k_x \lambda$, respectively, and the color bars on the top and bottom  panels indicate the time stages and the wave numbers, respectively.}
\label{fig:FIG2}
\end{figure*}
%----------------------------------------------------------------------------

Figure \ref{fig:FIG1} shows the time evolution of plasma density for $r_{\rm g}/\lambda=0.11$. Part of the simulation domain is depicted.  Three typical stages of the time evolution of reconnection are shown by the color scales indicated by the bars at the right-hand side of the panels. The white lines are magnetic field lines calculated from the contour of the vector potential $A_z(x,y)$.
The top panel is the early linear stage at $t/\tau_{\rm A}=290$, when the amplitude of the reconnecting magnetic field is gradually growing; the second panel is the nonlinear stage at $t/\tau_{\rm A}=519$, when thinning of the plasma sheet is clearly observed around the X-type point; and the third panel at $t/\tau_{\rm A}=533$ is the late nonlinear stage, when the elongation of the current sheet and the formation of several small plasmoids are observed.
The time evolution is basically same as many previous reconnection simulations \cite[e.g.,][]{Terasawa81,Hoshino87,Hesse01,Pritchett01,Hoshino01,Drake08}, but shown here is a thick current sheet in which the gyro-radius is much smaller than the thickness of the plasma sheet.

Note that other simulation runs for $r_g/\lambda=0.22$ and $0.45$ show almost the same time evolution as that of Figure \ref{fig:FIG1} (not shown here).  Namely, the plasma sheet gradually shrinks during the early time evolution, and after the formation of X-type neutral point at the center of the simulation box, the elongated thin current sheet is formed, where several small scale plasmoids are embedded.  We find the small scale plasmoids in the elongated current sheet has a tendency to be easily formed for the thick current sheet for $r_g/\lambda=0.11$.  

The top three panels in Figure \ref{fig:FIG2} show the time evolution of the power spectra for the Fourier components of $|A_z(k_x)|$ around $y=0$ as the function of wave number $k_x$.  $A_z(k_x)$ is obtained by
\begin{equation*}
  A_z(k_x) =\frac{1}{0.4 \lambda \pi} \int_{-0.1 \lambda}^{0.1 \lambda} A_z(x,y) \exp(i k_x x) dx dy.
\end{equation*}
From the left to the right-hand panel, the simulation results for $r_g/\lambda=0.11$, $0.22$, and $0.45$ are depicted, respectively.  The dark blue line represents the initial stage, the blue to green colors are the linear to the early nonlinear phase of the tearing mode/magnetic reconnection, and the red one is the nonlinear stage after the formation of the elongated thin current sheet.  The color bar on the top indicates the corresponding time stage.  We find that from the early time evolution with the dark blue to the green colors the vector potential $|A_z(k_x)|$ can grow only for $0 < k_x \lambda <1$, suggesting that the tearing mode is unstable for the modes with long wavelengths, and it has the maximum growth rate around $k_x \lambda=0.2 - 0.5$.

During the early nonlinear phase with the green to the yellow colors, the wave number with the maximum wave power of $|A_z(k_x)|$ is gradually shifted into the short wave number region.  The reddish lines correspond to the stage after the formation of the elongated thin current sheet around the X-type neutral point, and we find a spiky structure due to the elongated current sheet by analogy with the Fourier transformation for Heaviside function.  The wave number of the first spiky notch $k_{\rm notch}$ is related to the length of the elongated current sheet $l$ by $k_{\rm notch}= 2 \pi/l$, and its high harmonics appear at $k = 2 \pi n/l$ with $n=2,3,..$  The wave number $k_{\rm notch}$ is shifted to a small regime in association of the elongation of the current sheet.

The bottom three panels in Figure \ref{fig:FIG2} show the time histories of each Fourier components of $|A_z(k_x)|$ as the function of time $t/\tau_A$.  The data sets are same as the above three panels.  The color lines represent the wave number, which is indicated on the top color bar.  The reddish colors are the small wave numbers of $k_x \lambda \sim 0$, and the bluish colors are the large wave numbers of $k_x \lambda \sim 1$.  From the left to the right-hand panel, three simulation runs for $r_g/\lambda=0.11$, $0.22$, and $0.45$ are respectively shown.  If we focus on the time evolution of the Fourier mode with the maximum wave power shown by the yellowish/reddish color line, we find that there are two time stages of the linear growth phase and the rapid growth phase followed after the linear growth phase. The rapid growth phases appear around $t/\tau_A \sim 520$ (left), $230$ (middle), and $110$ (right), respectively.  In the following sections, we focus on the second phase with the rapid nonlinear growth.  Before discussing the rapid nonlinear phase, however, we analyze the linear phase obtained in our PIC simulations in the next section.
%----------------------------------------------------------------------------
\section{Linear Stage}
\begin{figure}
\includegraphics[width=6cm]{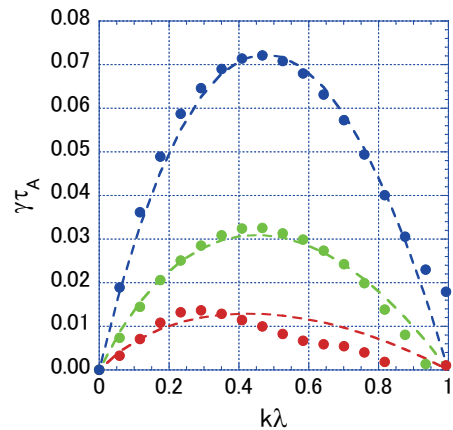}
\caption{Linear growth rates as functions of $k \lambda$ for three runs of $r_g/\lambda=0.11$ (red), $0.22$ (green), and $0.45$ (blue), respectively. The theoretical curves are depicted by the dashed lines.}
\label{fig:FIG3}
\end{figure}

%----------------------------------------------------------------------------
We shortly study the linear stage for the tearing instability/magnetic reconnection seen in Figure \ref{fig:FIG2}.
Shown in Figure \ref{fig:FIG3} is the linear growth rates calculated from the slopes of the time history of $|A_z(k_x)|$ in the bottom panels in Figure \ref{fig:FIG2} for three cases for $r_g/\lambda=0.11$ (red), $0.22$ (green), and $0.45$ (blue).  In Figure \ref{fig:FIG2}, the slopes of the wave modes with $k \lambda \sim 0$ and $k \lambda \sim 1$ are more or less fluctuating compared with other wave numbers around $k \lambda = 0.2 -0.5$, because the particle-in-cell simulation has intrinsic ``thermal'' noises due to a finite number of particles in the grid cell.  We made a least-square fitting for the slopes of $r_g/\lambda=0.11$, $0.22$, and $0.45$ by a function of $\exp(\gamma_L t)$ for the time interval of $t/\tau_{\rm A}=[10,150]$, $[10,70]$, and $[10,55]$, respectively.  The closed circles are the linear growth rate $\gamma_L$ obtained in the simulation runs.

The dashed curves superposed on the simulation data points are the theoretical linear growth rates of Equation (\ref{eq:modifiedlineargrowth}) in Appendix, which is given by
\begin{equation*}
  \gamma_{\rm l} \tau_{\rm A} =  \frac{\sqrt{\pi}}{2}  \left( \frac{r_{\rm g}}{\lambda} \right)^{3/2}
  \frac{(1 - k^2 \lambda^2) k \lambda}{k \lambda + \sqrt{r_g/\lambda}}.
\end{equation*}
As the inertia conductivity may be an order of estimation due to the straight orbit of the meandering particle (see Appendix), we have multiplied a factor of $0.73$ on the right-hand side of the above equation.  In addition, we assumed the background uniform plasma density $N_b$ in the initial condition, which may reduce the growth rate.
From Figure \ref{fig:FIG3},  
not only the dependence of the wave number $k \lambda$ but also the dependence of the the ratio of thickness $(r_g/\lambda)^{3/2}$  is consistent between the simulation results and the theory.  Note that the factor of $k \lambda/(k \lambda + \sqrt{r_g/\lambda})$ comes from the correction of the non-adiabatic energy (see Appendix). The theoretical growth rate is applicable only for $\sqrt{r_g/\lambda} < k \lambda <1$, nevertheless the consistency in the small $k\lambda < \sqrt{r_g/\lambda}$ is good.

%----------------------------------------------------------------------------
%----------------------------------------------------------------------------

Although we found the linear growth rates is basically consistent with the theoretical prediction, the profile for the thick current sheet with $r_g/\lambda = 0.11$ slightly differs from other two cases.  In addition, we also see some discrepancy between the theory and the simulation around $k \lambda \sim 1$.  As we have already mentioned, PIC simulations has a thermal fluctuation, which is proportional to $1/\sqrt{n_p}$, and it is not easy to demonstrate the linear growth phase for the wave modes with the relatively small growth rates.
%----------------------------------------------------------------------------
\begin{figure*}[t]
\includegraphics[width=18cm]{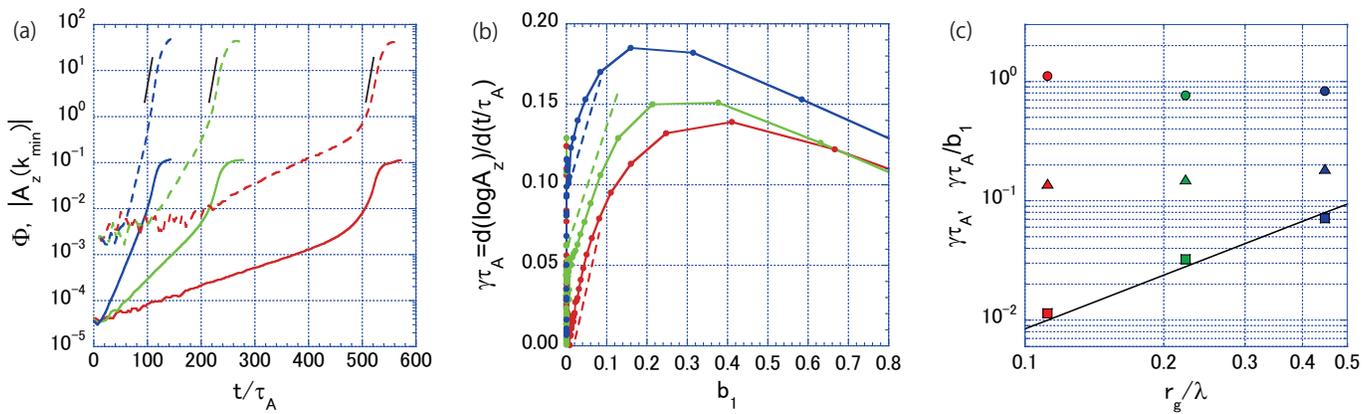}
\caption{(a) Time history of the reconnecting magnetic flux $\Phi$ (solid lines) and the Fourier component of $|A_z(k_{\rm min})|$ (dashed lines) as functions of time $t/\tau_{\rm A}$. The red, green, and blue lines correspond to the three cases of $r_{\rm g}/\lambda=0.11$, $0.22$, and $0.45$, respectively. The black solid lines with the same slope are shown as a reference.  (b) Instantaneous growth rate defined by $d\log A_y/d(t/\tau_{\rm A})$ as a function of the reconnecting magnetic field $b_1=B_y/B_0$.  Three dashed lines show the linear fitting of $\gamma \tau_A/b_1$ for the early nonlinear growing stage.  (c) The maximum nonlinear growth rates $\gamma \tau_A$ (triangles) and the slopes of the nonlinear growth rates $\gamma \tau_A/b_1$ (circles) as functions of $r_{\rm g}/\lambda$.  The linear growth rates $\gamma \tau_A$ (squares) and the linear theory of $(r_g/\lambda)^{3/2}$ (solid line) are shown as a reference.}
\label{fig:FIG4}
\end{figure*}
%----------------------------------------------------------------------------
\section{Non-linear Stage}
\subsection{PIC Simulation}
%----------------------------------------------------------------------------
To study the time evolution from the linear to the nonlinear stage, 
we show the time history of the Fourier component of $A_z(k_{\rm min})$ (dashed lines) and the reconnecting magnetic flux $\Phi$ (solid lines) in the same time scale in Figure \ref{fig:FIG4}a.  $\Phi$ is defined by
\begin{equation*}
  \Phi = \frac{1}{0.1 B_0 L_x \lambda}\int _{-0.1 \lambda}^{0.1 \lambda} \int_{0}^{L_x/2} B_y(x,y) dx dy,
\end{equation*}
and we choose the minimum wave number $k_{\rm min} \lambda =0.0584$ at the neutral sheet $y=0$.
Each set of three lines corresponds to the three different initial thicknesses of the plasma sheet, with $r_{\rm g}/\lambda=0.11$ (red), $0.22$ (green), and $0.45$ (blue), respectively.

As we have already mentioned that all evolution of $\Phi$ have two time scales: the first slopes observed in the stage of small $\Phi < 0.01$  corresponds to the linear growth phase given by $\Phi \propto \exp(\gamma_{\rm l} t)$.  The second slopes for $\Phi > 0.01$ show the nonlinear evolution of reconnection.  We find that the reconnection can rapidly grow after a certain stage of the linear tearing evolution.  All saturation levels for $\Phi$ are approximately $0.1$, which is consistent with the reconnection rate of $0.1$ \cite[e.g.,][]{Petschek64,Birn01}

The corresponding vector potentials $A_z$ depicted by the dashed lines show the same behavior as $\Phi$, but the slopes of $A_z$ are slightly steeper than those of $\Phi$.  While the linear growth rate of $A_z$ is a function of wave number $k$, the reconnecting magnetic flux of $\Phi$ is the integrated quantity in $k$-space.
Three black solid lines depicted as a reference in Figure \ref{fig:FIG4}a have the same slope in order to compare the time scale of the nonlinear growth rate.  The linear growth rate strongly depends on the thickness of the current sheet, but the early nonlinear growth rates are found to be almost independent of the current sheet thickness.

Let us investigate the growth rate in the nonlinear phase in details.  Shown in the middle panel of Figure \ref{fig:FIG4}b is the instantaneous growth rate defined by $d\log A_z/d(t/\tau_{\rm A})$ as the function of the reconnecting magnetic field $b_1 = B_y(k_{\rm min})/B_0=k_{\rm min} A_z(k_{\rm min})/B_0$ for the largest wavelength in the system with $k_{\rm min} \lambda = 0.0584$.  The linear growth phase should correspond to the region with small $b_1 \sim 0$, and we can confirm that these instantaneous growth rates are same as those obtained from Figure \ref{fig:FIG3}.  The reason why we investigate the instantaneous growth rate as the function of $b_1$ is to compare it with the nonlinear theory proposed by \citet{Galeev78}, which theoretical framework is discussed in the next subsection.

Focusing on the nonlinear phase,
we can see that the instantaneous growth rate increases with increasing amplitude $b_1$, and the maximum growth rates are about $\gamma \tau_A = 0.139$, $0.151$ and $1.85$ for $r_g/\lambda=0.11$, $0.22$, and $0.45$, respectively.   These maximum growth rates in the nonlinear phase are depicted in the right-hand panel of Figure \ref{fig:FIG4}c by the triangle symbols.  Three closed square symbols are the linear growth rates $\gamma \tau_A$ with the dependence of $(r_g/\lambda)^{3/2}$ for $k \lambda = 0.41$ in Figure \ref{fig:FIG3} as a reference. We find that the maximum growth rate is almost independent of the thickness of the current sheet $r_g/\lambda$.

More importantly, the growth rates evolve similarly in time and are linearly proportional to $b_1$.  The three circle symbols shown in Figure \ref{fig:FIG4}c are the measure of the nonlinear growth rates $\gamma \tau_A/b_1$ estimated from the slopes with the dashed lines in the middle panel of Figure \ref{fig:FIG4}b.   We find that both the maximum growth rates (triangles) and the slope of $d\log A_z/d(t/\tau_{\rm A})/b_1$ (circles) in the nonlinear phase do not depend on the initial thickness of the plasma sheet.

To define the amplitude of the reconnecting magnetic field $b_1$, we used the amplitude for the longest wavelength mode with $k_{\rm min} \lambda =0.0584$, but we may define $b_1$ for the maximum amplitude of $B_y(x)/B_0$ at the neutral sheet.  Even in this case, the slopes of $d\log A_z/d(t/\tau_{\rm A})/b_1$ are same among three cases, and do not depend on the initial thickness of the plasma sheet.

%----------------------------------------------------------------------------
\begin{figure}[b]
\includegraphics[width=6cm]{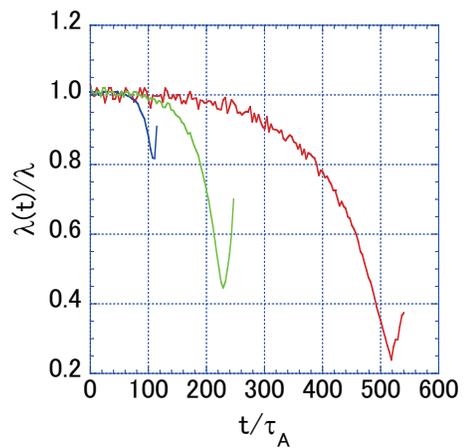}
\caption{Time histories of the thickness of the current sheet around the X-type neutral line $x/\lambda=54$.  The blue, green and red lines correspond to $r_{\rm g}/\lambda=0.45$, $0.22$, and $0.11$, respectively.}
\label{fig:FIG5}
\end{figure}
%--------------------------------------------------------------------------

It is also interesting to note the relationship between the nonlinear growing stage and the thickness of the current sheet, because thinning of the plasma sheet is observed in the middle panel of Figure \ref{fig:FIG1}.  
Shown in Figure \ref{fig:FIG5} is the time history of the thickness of the current sheet normalized by the initial thickness $\lambda$.  We fit the profile of the electric current in the $z$-component $J_z(y)$ around $x/\lambda=54$ by the function $\cosh^{-2}(y/\lambda(t))$.
Note that the initial current sheet profile can be described by $\cosh^{-2}(y/\lambda)$, and that the $\cosh$-type function fitting is a very good approximation during the thinning phase.

As we can see, the thickness decreases rapidly with increasing time, and the minimum thicknesses for $r_{\rm g}/\lambda=0.45$, $0.22$, and $0.11$ reach $\lambda(t)/\lambda \simeq 0.83$ (blue), $0.45$ (green), and $0.26$ (red), respectively.
Namely, we find that these thicknesses are almost twice the initial gyro-radii $r_{\rm g}$, suggesting that the thinning of the current sheet ceases after the incoming particles become unmagnetized from the magnetic flux transport.  Plasma sheet thinning has been observed by satellite observations in the Earth's magnetotail \citep[e.g.,][]{Asano03,Sharma08}.

More importantly, from Figures \ref{fig:FIG4}a and \ref{fig:FIG5}, we find that the timing of the minimum thickness of the current sheet coincides with the maximum slope appearing in the nonlinear stage.
For the case of $r_{\rm g}/\lambda =0.11$, this timing occurs at $t/\tau_{\rm A}=519$, which corresponds to the middle panel in Figure 1, and appears before the elongation of the current sheet around the X-type region.

The rapid shrinkage of the current sheet has been discussed by many people so far, and it has been argued in two-fluid framework including (electron) inertia term that the current sheet has a tendency to form finite-time singularity during reconnection, and as the result it drives fast and impulsive reconnection \citep[e.g.,][]{Ottaviani93,Cafaro98,Bhattacharjee05,Hirota15}.

%--------------------------------------------------------------------------
\subsection{Theoretical Interpretation}
%----------------------------------------------------------------------------
To understand the intriguing nonlinear reconnection growth mentioned above, we extend the idea of the nonlinear explosive reconnection proposed by \citet{Galeev78}.  As we reviewed the energy principle of reconnection in Appendix \cite{Schindler66,Biskamp71,Galeev78,Hoshino20}, the balance of the magnetic energy in the plasma sheet and the energy dissipated in the magnetic diffusion region can be given by
\begin{equation}
 \left( \frac{1}{2 k} \right) \left(\frac{1 - k^2 \lambda^2}{k \lambda^2} \right) =
 \frac{\omega_{\rm p}^2}{\sqrt{\pi} c^2} \gamma \frac{d_x}{v_{\rm th}} d_x d_y,
\label{eq:energy-principle}
\end{equation}
where $\gamma = d(\log A_1)/dt$. The left-hand side is the magnetic energy in the plasma sheet calculated from the adiabatic process \cite{White77}, and the right-hand side is the Ohmic dissipation in the diffusion region with size $d_x$ and  $d_y$ in the $x$- and $y$-direction, respectively.  We have used the inertia conductivity $(ne^2/m)(d_x/v_{\rm th})$ obtained in the collisionless tearing instability \cite{Coppi66,Hoh66,Hoshino20}.

%----------------------------------------------------------------------------
\begin{figure*}
\includegraphics[width=16cm]{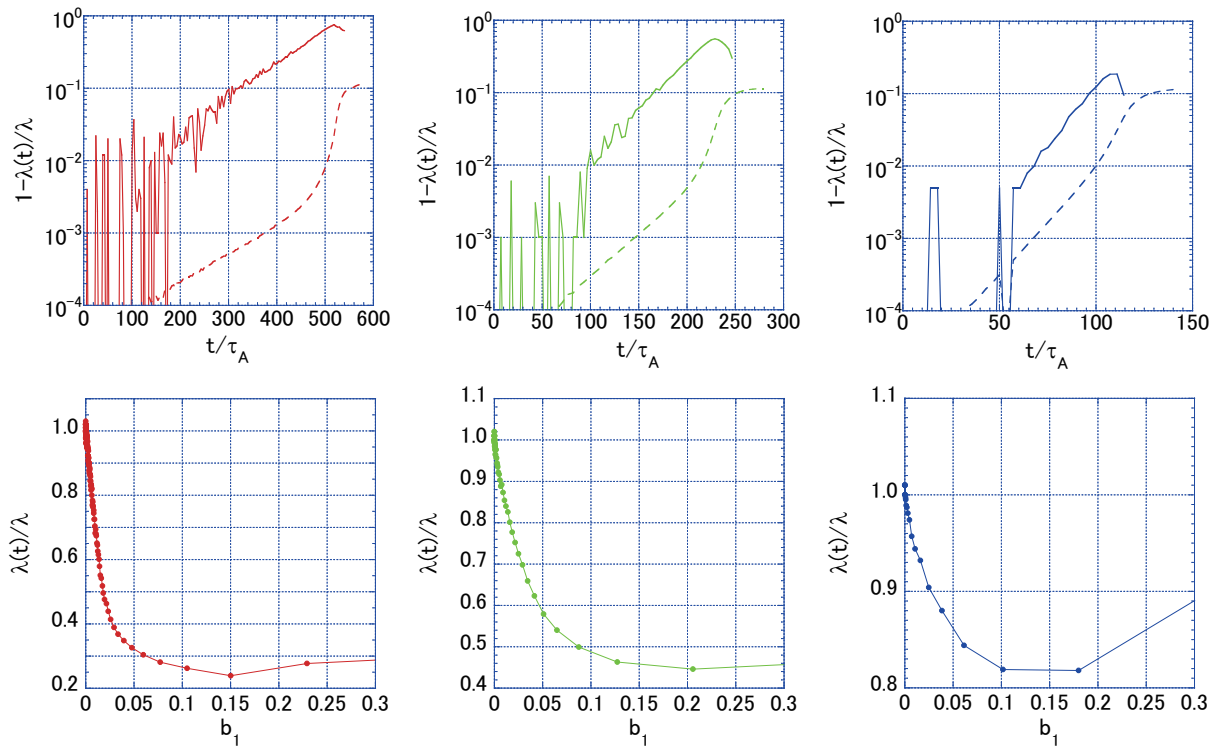}
\caption{Time histories of the thickness of the current sheet are shown for three cases of $r_{\rm g}/\lambda=0.11$ (red), $0.22$ (green), and $0.45$ (blue).  The top three panels are the measure of $1-\lambda(t)/\lambda$, and the time evolutions of the reconnecting magnetic flux $\Phi$ (dashed lines) are depicted as a reference. In the bottom panels, the corresponding time evolutions of the normalized thickness $\lambda(t)/\lambda$ are shown as a function of the reconnecting magnetic field $b_1$. }
\label{fig:FIG6}
\end{figure*}
%----------------------------------------------------------------------------

In the linear tearing instability \cite{Coppi66,Hoh66,Schindler66,Biskamp71,Galeev76}, $d_x$ is given by the wavelength of the tearing mode $k^{-1}$, and $d_y$ is the meandering width $d_y= \sqrt{\lambda r_{\rm g}}$. \cite[e.g.,][]{Sonnerup71}  Then the linear growth rate $\gamma_{\rm l}$ is obtained by
\begin{equation}
  \gamma_{\rm l} \tau_{\rm A} =  \frac{\sqrt{\pi}}{2} (1 - k^2 \lambda^2) \left( \frac{r_{\rm g}}{\lambda} \right)^{3/2}.
\end{equation}
We have assumed $d_y/\lambda < k \lambda <1$ and the constant $A_1(y)$ approximation.

In our comparison between the linear theory and the simulation in Figure \ref{fig:FIG3}, we included the correction term of a constant $A_1$ approximation, which is given by $k \lambda/(k \lambda + d_y/\lambda)$ with $d_y=\sqrt{r_g \lambda}$.  For simplicity, in the following discussion, however, we neglect the correction term of the constant $A_1$ approximation.  It would be desired to compare between theory and simulation for the long wavelength regime by keeping $\sqrt{r_g/\lambda} < k \lambda$, but it is very difficult to demonstrate such a thick current sheet in PIC simulation, because the linear growth rate is very slow.

We utilize this energy principle to explain the basic nature of the rapid growth of the nonlinear tearing mode.  As the reconnecting magnetic field $B_y$ increases, the non-adiabatic particle that participated in the Ohmic dissipation stars to be magnetized.  \citet{Galeev78} proposed that the size of $d_x$ is modified from the wave length of the tearing mode $k^{-1}$ to the local gyro-radius of $B_y$, which is given by $d_x = \sqrt{r_{\rm g}/(b_1 k)}$ if $b_1 > r_{\rm g} k$.
The nonlinear growth rate is given by
\begin{equation}
  \gamma_{\rm G} \tau_{\rm A} = \frac{\sqrt{\pi}}{2} \left( \frac{1 - k^2 \lambda^2}{k \lambda} \right)
      \left( \frac{r_{\rm g}}{\lambda} \right)^{1/2} b_1, 
\end{equation}
where a small reconnecting magnetic field $b_1 = B_y/B_0$.   The reconnecting magnetic field $B_y(x)$ at the neutral sheet is simply assumed to be $B_y(x) \sim B_y \tanh(kx)$.  The growth rate is proportional to the amplitude of the reconnecting magnetic field $b_1$, but this estimation includes the factor $(r_{\rm g}/\lambda)^{1/2}$.  This theoretical estimation is not consistent with our simulation result discussed in Figure \ref{fig:FIG4}c (triangles).
Note that the master equation of the energy principle is obtained in the linear approximation for the perturbed electric and magnetic fields, but the nonlinear effect emerges from the change of the area of the energy dissipation region.  

A key factor in explaining our simulation result is the thinning of the current sheet during reconnection as shown in Figure \ref{fig:FIG6}. The top three panels show the ``measure of the current sheet thickness'' defined by $1-\lambda(t)/\lambda$ for $r_{\rm g}/\lambda=0.11$ (red), $0.22$ (green), and $0.45$ (blue), respectively. As a reference, the time histories of the reconnecting magnetic fluxes $\Phi$ are depicted as dashed lines.  We find that the ``measure of the current sheet thickness'' grows exponentially in time, and that the shrinkage happens under the same growth rate as the linear growth rate $\gamma_{\rm l}$.  This result is consistent with the shrinkage of the current sheet studied in a two-fluid model by \citet{Hirota15}, who focused on the motion of the magnetic field line, i.e., the displacement of the contour $A_z={\rm const.}$ around the X-type neutral point, and found the exponential growth of the displacement.

The bottom three panels show the corresponding thickness $\lambda(t)$ as a function of $b_1$, and we find that the current sheets quickly shrink to the gyro-radius during the early nonlinear phase with $b_1 < 0.05$.  
In addition, by comparing the rate of shrinkage as a function of $b_1$ between the three cases, we find that the thick current sheet of $r_{\rm g}/\lambda=0.11$ (red) has the steepest slope of $\lambda(t)/b_1$, suggesting that the nonlinear explosive reconnection can easily happen for a thick current sheet.  Note that the explosive reconnection starts when
\begin{equation*}
  b_1 > r_g k,
\end{equation*}
and a thick current sheet with a small gyro-radius $r_g$ can initiate the nonlinear effect for a small reconnecting magnetic field $b_1=B_y/B_0$.

Because it is found that the current sheet can quickly shrink to the gyro-radius $r_{\rm g}$, we may estimate the nonlinear growth rate using $d_y = 2 r_{\rm g}$, obtaining
\begin{equation}
  \gamma_{\rm nl} \tau_{\rm A} =  \frac{\sqrt{\pi}}{4} \left( \frac{1 - k^2 \lambda^2}{k \lambda} \right) b_1.
\label{eq:nonlinear-growthrate}
\end{equation}
This nonlinear growth rate does not depend on $r_{\rm g}/\lambda$, and is consistent with our simulation result discussed in Figure \ref{fig:FIG4}.

We did not include the correction term of a non-constant $A_1$ approximation in the above discussion, but we may add the correction term of $k \lambda/(k \lambda + 2 r_g/\lambda)$ to Equation (\ref{eq:nonlinear-growthrate}) in order to argue the growth rate on the wave number dependence.  Then the modified growth rate can be given by
\begin{equation*}
  \gamma_{\rm nl} \tau_{\rm A} =  \frac{\sqrt{\pi}}{4} \left( \frac{1 - k^2 \lambda^2}{k \lambda+2 r_g/\lambda} \right) b_1,
\end{equation*}
and $\gamma_{\rm nl} \tau_{\rm A}/b_1$ become $2.33$, $1.41$, and $0.77$ for $r_g/\lambda=0.11$, $0.22$ and $0.45$, respectively.
Although the thoery gives only an order of estimation due to several simplified assumptions such as the stright meandering orbit and the twice of the gyro-radious for the shrinkage current sheet,
the theory is roughly consistent with the closed circles in Figure 4c.
Note that the theory is applicable only for the regime of $2r_g/\lambda < k \lambda < 1$, while the simulation data with $k \lambda=0.0584$ is obtained outside the above regime.  However, 
as we argued the linear growth rate as the function of the wave number $k \lambda$ beyond the assumption of $\sqrt{r_g/\lambda} < k \lambda <1$ in Figure \ref{fig:FIG3}, it might be useful to discuss the behavior in the short wave number regime.
As we can see the nonlinear evolution in Figure \ref{fig:FIG2}, the tearing mode with the smallest wave number can grow faster than others.
The behavior of the wave number dependence seems to be consistent with the above theoretical nonlinear growth rate.
%--------------------------------------------------------------------------
\section{Discussion and Summary}
%--------------------------------------------------------------------------
The debate regarding fast energy dissipation has remained a fundamental issue in magnetic reconnection for a thick plasma sheet, because the linear growth rate is known to decrease as the thickness of the current sheet is increased.  
In this paper, we have argued that the nonlinear process of reconnection in association with the local enhancement of the inertia resistivity leads to fast energy dissipation, the rate of which is independent of the initial thickness of the plasma sheet.
The local enhancement of the inertia resistivity around the X-type neutral point emerges by two effects of the shrinkage of the current sheet into the order of the particle gyro-radius and the magnetization of the non-adiabatic meandering particles by the reconnecting magnetic field $B_y$.  In the master equation of the energy principle in Equation (\ref{eq:energy-principle}), the former effect appears in $d_y$, and the latter one is $d_x$ due to the kinetic effect.

The explosive nature of reconnection has been already studied mainly in the framework of fluid approximation so far, and our finding of the shrinkage of initially thick current sheets in PIC simulations is basically same, but the enhancement of inertia resistivity proposed by \citet{Galeev78} has not been included in those numerical studies in the fluid approximation.  The constant nonlinear growth rate, which is independent of the initial thickness of current sheet, is realized by the help of the local enhancement of the inertia resistivity.

By integrating in time for both sides of Equation (\ref{eq:nonlinear-growthrate}) by noting $\gamma_{\rm nl}=d(\log b_1)/dt$, the amplitude increases explosively in time as
\begin{equation}
  b_1(t) \simeq \frac{b_1(0)}{1-(t-t_0)(b_1(0)\sqrt{\pi})/(4 k \lambda \tau_{\rm A})},
\end{equation}
where $b_1(0)$ is the amplitude at the start time $t_0$ of the nonlinear reconnection.
This equation has a singularity at
$\Delta t/\tau_A = (t -t_0)/\tau_A = (4 k \lambda)/(b_1(0) \sqrt{\pi})$.
As the coefficient of
$(4 k \lambda)/(b_1(0)\sqrt{\pi})$
is of the order of unity, the explosive nonlinear reconnection can occur in the Alfv\'{e}n transit time within a factor of the order of unity.  In fact, we find that the explosive growth in our simulation occurs in tens of the Alfv\'{e}n transit time.  During this explosive phase, however, the free energy on the left-hand side of Equation (\ref{eq:energy-principle}) is quickly exhausted, and thus the explosive reconnection may then cease.  The nonlinear saturation is beyond the scope of this paper.

We would like to mention the initial perturbation of $\delta A_z(x,y)$.  We discussed that the explosive reconnection occurs when $b_1 > r_g k$, and the critical amplitudes of $b_1 = B_y/B_0 = (r_g/\lambda) (k_{\rm min} \lambda)$ in our simulation can be given as $6.4 \times 10^{-3}$, $1.3 \times 10^{-2}$, and $2.6 \times 10^{-2}$ for $r_g/\lambda=0.11$, $0.22$ and $0.45$, respectively.  On the other hand, the initial perturbation in the simulation was set to be ${\rm max}(B_y/B_0)=10^{-3}$, which is smaller than the above critical amplitudes of the nonlinear reconnection.  This parameter setting makes possible to demonstrate the time evolution from the linear stage.
In fact, the initial X-type perturbation of $\delta A_z(x,y)$ has the maximum amplitude at $k_{\rm min} \lambda=0.0558$ in the Fourier space, which can be seen in the bottom purple/crimson lines for the top three panels in Figure 2, while the peak amplitudes in the linearly growing stage (blueish lines) are situated around $k \lambda= 0.2 \sim 0.5$.  And then, in the nonlinear stage, the most unstable mode is shifted toward the longest wavelength.
This confirms that the initial perturbation was set to be small so as to discuss the separation of two time stages of the linear and the nonlinear evolution. 

We investigated the nonlinear process to drive a fast magnetic reconnection for a pair plasma, but we think that this nonlinear process is applicable to the case for an ion-electron plasma as well.  During the early time evolution, the meandering electron around the X-type neutral point first starts to be magnetized, and if the electron scale current sheet embedded in the ion scale current is formed, it would be possible to observe the explosive reconnection in the electron scale.  During the later time evolution, the major player of the energy dissipation will be switched to the ion inertia resistivity, and we may observed it in the ion scale as well.  However, if the current sheet thickness becomes of the order of ion inertia length, the Hall current and the ambipolar electrostatic field become important \citep{Hoshino87,Birn01,Drake06}.  The electron stabilization effect due to the magnetized electrons is another open question \citep{Lembege82,Pellat91,Sitnov10}.

It would be important to mention the guide field $B_z$ effect.  We discussed the collisionless reconnection for an anti-parallel magnetic field topology without $B_z$, in which the inertia resistivity can be provided by unmagnetized particles.  For the guide field case, however, the inertia resistivity can be supplied by magnetized particles moving along the magnetic field line \citep{Drake77}.   Therefore, our claim of  the enhancement of the inertia resistivity due to the magnetization effect of the reconnecting magnetic field will not occur for a strong guide field case.

Our final comment is the nonlinear stage.  In the late nonlinear stage after the thinning of the plasma sheet, we observed the formation of the elongated current sheet \cite[e.g.,][]{Ottaviani93,Waelbroeck93,Jemella03,Loureiro05} in association with several small-scale plasmoids in Figure \ref{fig:FIG1}. Although the stage appears after the saturation of the reconnecting flux, it can lead to dynamic evolution of reconnection by launching magnetic islands and turbulent waves, and this stage is another important phase of reconnection discussed by many papers \cite[e.g.][]{Loureiro05,Drake06,Daughton07,Samtaney09,Hoshino12b,Higashimori13,Sironi14,Hoshino15}.  Our claim of the explosive reconnection can bridge the gap between the late nonlinear phase and the linear phase.
%--------------------------------------------------------------------------
\acknowledgments
  This work was supported by JSPS KAKENHI Grant Number 19H01949 and 20K20908.
%--------------------------------------------------------------------------
\section*{DATA AVAILABILITY}
The data support the findings of this study are available from the corresponding author upon reasonable request.
%--------------------------------------------------------------------------
\appendix
\section{Linear Theory of Collisionless Tearing Mode}
%--------------------------------------------------------------------------
Here, we review the basic framework of the energy principle of the tearing mode instability \cite[e.g.,][]{Coppi66,Hoh66,Galeev76,Biskamp00,Schindler10}.  We study the plasma sheet with the magnetic field $B_x = B_0 \tanh(y/\lambda)$.  For the perturbed physical quantities, the magnetic energy conservation integrated over the system can be written as
\begin{eqnarray}
  \frac{\partial}{\partial t} \iint \frac{B_1^2}{8 \pi} dx dy= -\iint E_1 J_1 dx dy,
\label{eq:energyprinciple1}
\end{eqnarray}
under a two-dimensional assumption.  The Poynting flux term $\iint \nabla \cdot (E_1 \times B_1) dx dy$ disappears due to the periodic boundary condition in the $x$-direction and $E_1 = 0$ at the conducting boundary at $y=0$ and $y=L_y$.
The $x$- and $y$-components of the magnetic field $B_1$ and the $z$-component of the electric field $E_1$ and $J_1$ are considered.  
The perturbed electric current $J_1$ consists of $J_1^{\rm ad}$ of the adiabatic change of the diamagnetic plasma current and $J_1^{\textrm non-ad}$ of the non-adiabatic current provided by the Landau resonance between the inductive reconnection electric field and the meandering particle.  We can thus rewrite the above equation as
\begin{equation}
 \iint  \left( \frac{\partial}{\partial t} \left( \frac{B_1^2}{8 \pi} \right)  + E_1 J_1^{\rm ad} \right) dx dy 
  =  - \iint E_1 J_1^{\textrm non-ad} dx dy.
\label{eq:energyprinciple2}
\end{equation}

For the tearing-mode instability, the Ohmic dissipation/heating is negligible except for the central current sheet, and the external solution of the vector potential $A_1(x,y,t)=A_1(y) \exp(ikx-i\omega t)$ describing the formation of the tearing island can be given by \cite{White77}
\begin{equation}
  A_1(y)= A_1(0) \left(1+\frac{\tanh (|y|/\lambda)}{k \lambda} \right) \exp(- k |y|),
\label{eq:externalsolution}
\end{equation}
where $A_1$ is the $z$-component, $B_{1}= \nabla \times A_1$ and $\lambda$ is the thickness of the plasma sheet.
After substituting Equation (\ref{eq:externalsolution}) into the left-hand side of Equation (\ref{eq:energyprinciple2}) using the relationships $E_1=-(1/c) \partial A_1 /\partial t$ and $J_1^{\rm ad} = (c/4 \pi) \nabla \times (\nabla \times A_1)$, 
and we integrate by parts, we obtain,
\begin{eqnarray}
  & & \int_{-\pi/k}^{\pi/k} \int_{-\infty}^{\infty}  \left( \frac{\partial}{\partial t} \left( \frac{B_1^2}{8 \pi} \right)  + E_1 J_1^{\rm ad} \right) dx dy \nonumber \\
  &=& \frac{1}{8 \pi} \frac{\partial}{\partial t} \iint \left( |\frac{d A_1}{dy} |^2 + k^2 |A_1|^2
  - \frac{2 |A_1|^2}{\lambda^2 \cosh^2(y/\lambda)} \right) dx dy  \nonumber \\
  &=& \frac{1}{4 k} \frac{\partial}{\partial t} \left( \Big| \frac{d A_1(y,t)}{dy} A_1(y,t) \Big|_{y=0}^{\infty} \right) \nonumber \\
  &=& \frac{\partial}{\partial t} \left( \frac{1}{4 k} \frac{(k^2 \lambda^2-1)}{k \lambda^2} |A_1(0,t)|^2 \right).
\label{eq:ad-energy}
\end{eqnarray}
Note that $A_1 d^2A_1/dy^2 = k^2 A_1^2 - 2 A_1^2/\lambda^2 \cosh^2(y/\lambda)$.

The non-adiabatic current can be calculated from the linearized Vlasov equation (see Appendix B), and we estimate it using Ohm's law and the collisionless inertia conductivity given by Equation (\ref{eq:inertia_conductivity}), i.e., 
$\sigma=2 \sqrt{\pi} n e^2/(m \nu)$
for a pair plasma.  Here, $\nu$ is the effective collision frequency estimated by the transit time of the thermal particle across the magnetic diffusion region, i.e., $\nu = v_{\rm th}/d_x$, where $d_x$ is the size of the diffusion region in the $x$-direction \cite[e.g.,][]{Coppi66,Hoh66,Galeev76,Biskamp00,Schindler10}.

The magnetic energy change provided by the Ohmic dissipation for the right-hand side term in Equation (\ref{eq:energyprinciple2}) can be written as
\begin{eqnarray}
  - \int_{-d_x}^{d_x} \int_{-d_y}^{d_y}  E_1 J_1^{\textrm non-ad} dx dy = - 4 \sigma |E_1|^2 d_x d_y \nonumber \\
  = -\frac{1}{\sqrt{\pi}} \frac{\omega_{\rm p}^2}{c^2}\frac{d_x}{v_{\rm th}} d_x d_y  \Big|\frac{\partial A_1(0,t)}{\partial t} \Big|^2,
\label{eq:nonad-energy}
\end{eqnarray}
where $\omega_{\rm p} = \sqrt{8 \pi n e^2/m}$ is the plasma frequency, and $d_x$ and $d_y$ are the lengths of the diffusion region in the $x$- and $y$-direction, respectively.  We have assumed the so-called constant $A_1$ approximation in the diffusion region, i.e., $A_1(y) \simeq A_1(0)$ for $|y| \leq d_y$.

From Equations (\ref{eq:ad-energy}) and (\ref{eq:nonad-energy}), we obtain
\begin{eqnarray}
  \left( \frac{1}{2 k} \right) \left( \frac{1-k^2 \lambda^2}{k \lambda^2}  \right)
  = \frac{\omega_{\rm p}^2}{\sqrt{\pi} c^2} \gamma \frac{d_x}{v_{\rm th}} d_x d_y,
\label{eq:energy-principle2}
\end{eqnarray}
where $\gamma = d(\log A_1)/dt$ is the growth rate.
The left-hand side is the magnetic energy in the plasma sheet calculated from the adiabatic process \cite{White77}, and the right-hand side is the Ohmic dissipation in the diffusion region with size $d_x$ and  $d_y$ in the $x$- and $y$-direction, respectively.  This is a master equation to analyze the linear and nonlinear explosive magnetic reconnection.

In the linear tearing instability \cite{Coppi66,Hoh66,Schindler66,Biskamp71,Galeev76}, $d_x$ is given by the wavelength of the tearing mode $k^{-1}$, and $d_y$ is the meandering width $d_y= \sqrt{\lambda r_{\rm g}}$. \cite[e.g.,][]{Sonnerup71}  The linear growth rate $\gamma_{\rm l}$ is obtained by
\begin{equation}
  \gamma_{\rm l} \tau_{\rm A} =  \frac{\sqrt{\pi}}{2} (1 - k^2 \lambda^2) \left( \frac{r_{\rm g}}{\lambda} \right)^{3/2}.
\label{eq:standardlineargrowth}
\end{equation}

In deriving the above linear growth rate, we have assumed the constant $A_1(y)=A_1(0)$ approximation in the diffusion region, but the profile of the vector potential $A_1(y)$ is not necessarily constant for a small $k \lambda \ll 1$.   An inner solution of $A_1^{\rm in}(y)$ in the diffusion region should have a concave form, which should be connected to the external solution of Equation (\ref{eq:externalsolution}) at $y=d_y$.  Here, for simplicity, we estimate the non-adiabatic energy change in the diffusion region by using the external solution of $A_1(y)$, and we obtain
\begin{equation}
  \int_{0}^{d_y} A_1^{\rm in}(y)^2 dy \sim \frac{d_y}{k \lambda} \left(k \lambda + \frac{d_y}{\lambda} \right) A_1(0)^2.
\end{equation}
Note that we have assumed $k d_y < k \lambda < 1$.
Therefore, the linear growth rate with a better approximation beyond the constant $A_1$ can be given by
\begin{equation}
  \gamma_{\rm l} \tau_{\rm A} =  \frac{\sqrt{\pi}}{2} \left( \frac{r_{\rm g}}{\lambda} \right)^{3/2}
  \frac{(1 - k^2 \lambda^2) k \lambda}{k \lambda + \sqrt{r_g/\lambda}}.
\label{eq:modifiedlineargrowth}
\end{equation}
Note that, strictly speaking, this linear growth rate is only applicable to $\sqrt{r_g/\lambda} < k \lambda < 1$.
%--------------------------------------------------------------------------
\section{Inertia Conductivity in Collisionless Tearing Mode}
We review the idea of the inertia conductivity used in the collisionless tearing mode.  According to the standard procedure of the tearing mode theory \citep[e.g.][]{Coppi66,Hoh66,Galeev76,Hoshino20},
the linearized Vlasov equation for a 2D tearing mode under the Harris distribution function in the $x-y$ plane can be given by
\begin{eqnarray}
\left( \frac{\partial}{\partial t} + 
  {\bf v} \cdot \frac{\partial}{\partial {\bf x}} + 
  \frac{e_j}{m c}({\bf v} \times {\bf B}_0) \cdot \frac{\partial}{\partial {\bf v}} \right) f_{1j}  \nonumber \\
 = \frac{e_j f_{0j}}{c T_j} 
  \left(  (\frac{\partial}{\partial t} + {\bf v} \cdot \nabla) u_j A_{1z}
  -\frac{\partial}{\partial t}(v_z A_{1z}) \right),
  \label{eq:linearizedVlasov}
\end{eqnarray}
where $f_{j}({\bf x},{\bf v},t)$ is the velocity distribution function for a species $j$, $u_j$ is the drift velocity, and the magnetic field ${\bf B} = \nabla \times {\bf A}$.  The suffixes of $0$ and $1$ represent the zero-th and the first order quantities, respectively.
We have assumed that $A_{1x}$ and $A_{1y}$ are small compared to $ A_{1z}$.
The zero-th order distribution function of the Harris equilibrium can be given by,
\begin{equation*}
  f_{0j} = n(y) \left( \frac{m_j}{2 \pi T_j} \right)^{3/2}
  \exp \left(- \frac{m_j}{2 T_j} (v_x^2 + v_y^2 + (v_z -u_j)^2) \right).
\end{equation*}

After the time integration of Equation (\ref{eq:linearizedVlasov}) along unperturbed particle trajectories from $-\infty$ to the current time $t$, we obtained
\begin{equation}
f_{1j} = \frac{e_j f_{0j}}{c T_j} \left(  u_j A_{1z}- \int_{-\infty}^{t} \frac{\partial}{\partial t} (v_z A_{1z}) dt' \right). 
\label{eq:tearingF}
\end{equation}
The first term without the symbol of the time integration in Equation (\ref{eq:tearingF}) is referred to as the adiabatic/ideal MHD term, which represents the contribution from the magnetized motion of particles resulting from the slow plasma convection motion. The second term involving the symbol of the time integration is identified as the non-adiabatic/non-ideal-MHD term, which represents the contribution from the departure from the foregoing adiabatic motion \citep[e.g.][]{Hoshino87}. 

Let us focus on the non-adiabatic term. As noting that the perturbation is represented by the frequency $\omega$ and the wave number $k$, and $A_{1z}(x,y,t)=A_{1z}(y) \exp(ikx-i \omega t)$, we obtain
\begin{equation}
f_{1j}^{\rm non-ad}
= \frac{e_j f_{0j}}{c T_j}  i\omega \int_{-\infty}^{t} v_z A_{1z}(y') e^{ikx'-i\omega t'} dt'. 
\end{equation}
The main contribution comes from the meandering particle orbit near the neutral sheet, whose thickness $d_j=\sqrt{r_g \lambda}$.  As in the conventional assumption of the particle trajectory motion, we may assume that the meandering motion can be treated as a type of free motion with a straight orbit,
i.e., $x'-x=v_x(t'-t)$ and $y'=y$.  Thereafter, the time integration of the above equation becomes
\begin{eqnarray}
f_{1j}^{\rm non-ad} &\simeq& \frac{e_j f_{0j}}{c T_j} i\omega \int_{-\infty}^{t} 
v_z A_{1z}(y) e^{ikx-i\omega t} e^{i(kv_x-\omega)(t'-t)} dt' ,  \nonumber \\
&\simeq& \frac{e_j f_{0j}}{c T_j} i\omega v_z A_{1z}(y) e^{ikx-i\omega t}
\int_{-\infty}^{0} e^{i(kv_x-\omega) \tau} d \tau ,  \nonumber \\
& \simeq & \frac{e_j f_{0j}}{c T_j} \frac{\omega}{\omega - kv_x + i0} v_z A_{1z} H(d_j),
\label{eq:tearingF2}
\end{eqnarray}
where $H(l_j)$ is a Heaviside function with $H(d_j)=1$ for $|y| \le l_j$ or $H(d_j)=0$. The singularity of the denominator shows the Landau resonance between the electric field $E_{1z}=-(1/c)(\partial A_{1z}/\partial t)$ and the meandering particle 
for the tearing mode with the zero frequency wave of $\rm{Re}(\omega)=0$. 

We can now obtain the relationship between the electric current coming from the non-adiabatic motion and the inductive electric field around the neutral sheet. By substituting the Harris equilibrium distribution function, we obtain
\begin{eqnarray}
J^{\rm non-ad}_{1z}
&\simeq& \sum_{\pm} \int e_j v_z
\frac{e_j f_{0j}}{c T_j} \frac{\omega v_z}{\omega - kv_x + i0} A_{1z} d^3v, \nonumber \\
&\simeq& -i\sum_{\pm} \frac{n e^2}{m_j} \frac{1}{k v_{th,j}}
       (1+ \frac{m_j u_j^2}{T_j})Z(\frac{\omega}{k v_{th,j}}) E_{1z},  \nonumber \\
&\simeq& \sum_{\pm} \frac{n e^2}{m_j} \frac{\sqrt{\pi}}{k v_{th,j}} E_{1z},
\end{eqnarray}
where $Z(\omega/k v_{th,j})$ is the plasma dispersion function.  
We have assumed that the drift velocity $u_j$ and $\omega/k$ are smaller than the thermal velocity $v_{th,j}$.
Therefore, the inertia conductivity $\sigma=J^{\rm non-ad}/E$ for a pair plasma can be expressed as,
\begin{equation}
  \sigma \simeq \frac{2 ne^2}{m} \frac{\sqrt{\pi}}{k v_{th}}.
\label{eq:inertia_conductivity}
\end{equation}
This equation is consistent with the idea such that the effective collision frequency $\nu$ is given by $k v_{th}$ and that the meandering particle traveling along the x-axis with the thermal velocity $v_{th}$ can be scattered by the tearing island with the scale length of $k^{-1}$.
%--------------------------------------------------------------------------
%\newpage

%----------------------------------------------------------------------------
\end{document}